\documentclass[twocolumn, switch]{article} 

\usepackage{preprint}

\usepackage{amsmath, amsthm, amssymb, amsfonts}

\usepackage[numbers,square]{natbib}
\usepackage[utf8]{inputenc}	
\usepackage[T1]{fontenc}	
\usepackage{xcolor}		
\usepackage[colorlinks = true,
            linkcolor = purple,
            urlcolor  = blue,
            citecolor = cyan,
            anchorcolor = black]{hyperref}	
\usepackage{booktabs} 		
\usepackage{nicefrac}		
\usepackage{microtype}		
\usepackage{lineno}		
\usepackage{float}			

\usepackage{lipsum}		

\usepackage{newfloat}
\DeclareFloatingEnvironment[name={Supplementary Figure}]{suppfigure}
\usepackage{sidecap}
\sidecaptionvpos{figure}{c}

\usepackage{titlesec}
\titlespacing\section{0pt}{12pt plus 3pt minus 3pt}{1pt plus 1pt minus 1pt}
\titlespacing\subsection{0pt}{10pt plus 3pt minus 3pt}{1pt plus 1pt minus 1pt}
\titlespacing\subsubsection{0pt}{8pt plus 3pt minus 3pt}{1pt plus 1pt minus 1pt}

\usepackage{array}
\usepackage{multirow}
\usepackage[]{threeparttable}

\newcolumntype{P}[1]{>{\centering\arraybackslash}p{#1}}

\title{Learning effective physical laws for generating cosmological hydrodynamics with Lagrangian Deep Learning}

\usepackage{xwatermark}
\newwatermark[firstpage,color=gray!90,angle=0,scale=0.28, xpos=0in,ypos=-5in]{*correspondence: \texttt{biwei@berkeley.edu}}

\usepackage{authblk}

\author[a\thanks{\tt{biwei@berkeley.edu}}]{Biwei Dai}
\author[a,b,c]{Uro{\v s} Seljak}
\affil[a]{Berkeley Center for Cosmological Physics and Department of Physics, University of California, Berkeley, CA 94720, USA}
\affil[b]{Department of Astronomy, University of California, Berkeley, CA 94720, USA}
\affil[c]{Lawrence Berkeley National Lab, 1 Cyclotron Road, Berkeley, CA 94720, USA}

\begin{document}

\twocolumn[ 
  \begin{@twocolumnfalse} 
  
\maketitle

\begin{abstract}
The goal of generative models is to learn the intricate relations between the data to create new simulated data, but current approaches fail in very high dimensions. When the true data generating process is based on physical processes these impose symmetries and constraints, and the generative model can be created by learning an effective description of the underlying physics, which enables scaling of the generative model to very high dimensions. 
In this work we propose Lagrangian Deep Learning (LDL) for this purpose, applying it to learn outputs of cosmological hydrodynamical simulations. 
The model uses layers of Lagrangian displacements of particles describing the observables to learn the effective physical laws. The displacements are modeled as the gradient of an effective potential, which explicitly satisfies the translational and rotational invariance. 
The total number of learned parameters is only of order 10, and they can be viewed as effective theory parameters.
We combine N-body solver FastPM with LDL and apply them to a wide range of cosmological outputs, from the dark matter to the stellar maps, gas density and temperature. The computational cost of LDL is nearly four orders of magnitude lower than the full hydrodynamical simulations, yet it outperforms it at the same resolution. We achieve this with only of order 10 layers from the initial conditions to the final output, in contrast to typical cosmological simulations with thousands of time steps. 
This opens up the possibility of analyzing cosmological observations entirely within this framework, without
the need for large dark-matter simulations.
\end{abstract}
\keywords{deep learning \and Lagrangian approach \and cosmological hydrodynamical simulation} 
\vspace{0.35cm}

  \end{@twocolumnfalse} 
] 



\section{Introduction}
Numerical simulations of large scale structure formation in the universe are essential for extracting cosmological information from the observations \cite{eisenstein2011sdss, ivezic2009lsst, spergel2015wide, aghamousa2016desi, amendola2018cosmology, springel2005simulations, derose2019aemulus}. In principle hydrodynamical simulations are capable of predicting the distribution of all observables in the universe, and thus can model observations directly. However, running high resolution hydrodynamical simulations at volumes comparable to the current and future sky surveys is currently not feasible, due to its high computational costs. The most widely used method is running gravity-only N-body simulations, and then populating baryons in the halo catalogs with semi-analytical approaches such as halo occupation 
distribution (HOD) \cite{berlind2002halo}, or halo assembly models \cite{behroozi2019universemachine}. 
However, these methods make strong assumptions such as the halo mass being the main quantity controlling the baryonic properties. 
In addition, many of the cosmological observations such as X-ray emission and 
Sunyaev-Zeldovich emission are based 
on hydrodynamic gas properties such as gas density, 
temperature, pressure etc., which cannot be 
modeled in the dark matter only simulations. 

Deep learning methods provide an alternative way to model the cosmological observables. A number of papers view the task as an image-to-image translation problem, i.e., they take in pixelized matter density field as input data, and output the target pixelized observable field. These methods either model the conditional probability distribution $p(y_{\mathrm target}|x_{\mathrm input})$ using deep generative models such as GANs \cite{goodfellow2014generative} and VAEs \cite{kingma2013auto, rezende2014stochastic}, or learn a mapping $x_{\mathrm input} \mapsto y_{\mathrm target}$ with deep convolutional neural networks (DCNN). Previous work in this area covers a wide array of tasks, such as identifying halos (protohalos) \cite{modi2018cosmological, berger2019volumetric, ramanah2019painting, bernardini2020predicting}, producing 3D galaxy distribution \cite{zhang2019from}, generating tSZ signals \cite{troster2019painting}, predicting dark matter annihilation feedback \cite{list2019black}, learning neutrino effects \cite{giusarma2019learning}, emulating high resolution features from low resolution simulations \cite{ramanah2020super} etc.

Unlike these methods that work in pixel (Eulerian) space and treat the field as images, another way to model the dynamics is to adopt the Lagrangian scheme, i.e., trace the motion of individual particles or fluid elements by modeling their displacement field. The displacement field contains more information than the density field, as different displacement fields can produce the same density field, and is in general more Gaussian and linear than the density field. Existing 
methods in 
this space only cover the dark matter, e.g. approximate N-body solvers \cite{tassev2013solving, feng2016fastpm} and DCNN \cite{he2019learning}.

In this work we propose a novel deep learning architecture, Lagrangian Deep Learning (LDL), for modeling both the cosmological dark matter and the
hydrodynamics, using the Lagrangian approach. The model is motivated by the effective theory ideas in 
physics, where one describes the true process, 
which may be too complicated to model, 
with an effective, often coarse grained, description of physics. A typical example 
is the effective field theory, where perturbative 
field theory is supplemented with an effective 
field theory terms that obey the symmetries, 
and are an effective coarse-grained 
description of the non-perturbative small 
scale effects. 
The resulting effective description 
has a similar structure as the true physics, but 
with free coefficients that one must fit 
for, and that account for the non-perturbative small 
scale effects. 

\section*{Lagrangian Deep Learning}

A cosmological dark matter and baryon evolution can be described by a system of partial 
differential equations (PDE) coupling gravity, hydrodynamics, and various sub-grid physics
modeling processes such as the
star formation, which are evolved in 
time from the beginning of the universe until today. 
One would like to simulate a 
significant fraction of the observable 
universe, while also capturing important physical 
processes on orders of magnitude
smaller scales, all in three dimensions. As a result
the resulting dynamical range is excessive even for the
modern computational platforms. As an 
example, the state of the art Illustris TNG300-1 \cite{pillepich2018first, naiman2018first, marinacci2018first, springel2018first, nelson2018first}
has of order $10^{10}$ particles, yet 
simulates only 
a very small fraction of observable 
universe. The lower resolution TNG300-3
reduces the number of particles by 64, at a 
cost of significantly reducing 
the realism of the simulation. 

An effective physics approach is to rewrite 
the full problem into a large scale problem that 
we can solve, together with an 
effective description of the small scales which 
we cannot resolve. In theoretical physics 
this is typically done by rewriting the 
Lagrangian such that it takes the most 
general form that satisfy the symmetries 
of the problem, with free coefficients 
describing the effect of the small scale coarse-graining. In cosmology the large scale
evolution is governed by gravity, which 
can easily be solved perturbatively or 
numerically. Effective descriptions using 
perturbative expansions exist \cite{carrasco2012effective}, but 
fail to model small scales and complicated 
baryonic processes at the map level. 
While spatial coarse graining is the 
most popular implementation of this idea, 
one can also apply it to temporal 
coarse graining as well. A typical PDE 
solver requires many time steps, which is 
expensive. 
Temporal coarse graining replaces this with 
fewer integration time steps, at a price of replacing 
the true physics equations with their effective 
description, while ensuring the true 
solution on large scales, where the 
solution is known \cite{tassev2013solving,feng2016fastpm}. 

Here we take this effective physics description idea and combine it with the deep learning paradigm, 
where one maps the data through 
several layers 
consisting of simple operations, and 
trains the coefficients of these 
layers on some loss function of choice. While 
machine learning layers are 
described with neural networks with a very
large number of coefficients, here we will
view a single layer as a single time step
PDE solver, using a similar structure as the 
true physical laws. This has the advantage 
that it can preserve the symmetries inherent 
in the problem. The main symmetry we wish 
to preserve in a cosmological setting is the
translational and rotational symmetry: the physical laws have no preferred position or 
direction. But we also wish to satisfy the existing conservation laws, such as the 
dark matter and 
baryon mass conservation. 

A very simple implementation of these two 
requirements is Lagrangian displacements
of particles describing the dark matter or
baryons.  
We displace the particles using the gradient of a potential, and mass 
conservation is ensured since we only 
move the particles around. 
To ensure 
the translation and rotation symmetry within
the effective description we shape 
the potential in Fourier space, such that it 
only depends on the amplitude of the Fourier 
wave vector. 
The potential gradient can be viewed as a
force acting upon 
their acceleration via the Newton's law, 
and the shaping of the potential is equivalent 
to the radial dependence of the force. 
This description requires particle positions 
and velocities, so it is a second order PDE in time. 
We will use this description for the dark 
matter. However, for baryons we can simplify 
the modeling by assuming their velocity 
is the same as that of the dark matter, since
velocity is dominated by large scales where the two trace each other. 
In this case we can use the 
potential gradient to displace particle
positions 
directly, so the description becomes 
effectively first order in time. Moreover, 
by a simple extension of the model we can apply this concept to the 
baryonic observables such as the gas pressure
and temperature, where conservation 
laws no longer apply. A complete description 
also requires us to define the source for the 
potential. In physics this is typically 
some property of the particles, such as mass
or charge. Here we wish to describe the 
complicated nonlinear processes of 
subgrid physics, as well as coarse graining in 
space and time. Motivated by gravity 
we will make the simplest 
possible assumption of the source being a simple
power law of the density, using a learned Green's function to convert to the potential. 
Since we wish to model several different 
physics processes we stack it into 
multiple layers. 
Because the model takes in the particle data and models the displacement field from the Lagrangian approach using multiple layers, we call this model Lagrangian Deep Learning (LDL).

Our specific goal is to model the distribution of dark matter and hydrodynamic observables 
starting from the initial conditions as 
set in the early universe, using an effective 
description that captures the physics 
symmetries and conservation laws. 
An 
example of such a process applied to 
time and spatial coarse graining is the 
dark matter evolution with a few time steps 
only, which combines ideas such as the
approximate N-body solvers, with 
a force sharpening process called the Potential Gradient Descent (PGD) to capture the
coarse graining \cite{dai2018gradient,dai2020high}. 
We first use FastPM \cite{feng2016fastpm}, a quasi particle-mesh (PM) N-body solver, which ensures the correct large scale growth at any number of time steps, since the kick and drift factors of the leapfrog integrator in FastPM are modified following the linear  (Zel’dovich) equation of motion. 
FastPM has a few layers only (typically 5-10)
and uses particle displacements. 
It is supplemented by one additional layer
of PGD applied to position only 
to improve the dark matter distribution 
on small 
scales.  All of the steps of this process
are in the LDL form, so can be viewed as 
its initial layers. 
The resulting dark matter maps are 
shown in figure \ref{fig:visual} and show an excellent 
agreement with the full N-body simulation of 
Illustris TNG, which is also confirmed by 
numerical comparisons presented in \cite{dai2018gradient}. This application is not 
learning new physics, but is learning the effective 
physics description 
of both time and spatial coarse graining: 
instead of 1000+ time steps in a standard N-body 
simulation we use only 10, and instead of the 
full spatial resolution we will use a factor of 64 
reduced mass resolution. 

Here we wish to extend these ideas to the 
more complex and expensive problem of cosmological 
hydrodynamics, where we wish to learn 
its physics using an effective description. Baryons are 
dissipative and collisional, with many 
physical processes, such as cooling, 
radiation, star formation, gas shocks, turbulence etc. happening inside the 
highest density regions called dark matter halos. 
One can add displacements to the dark matter particles to simulate these hydrodynamic processes, such that the particles after the displacement have a similar distribution as the baryons. Enthalpy Gradient Descent (EGD) is an example of this idea \cite{dai2018gradient}: one adds small scale displacement to the dark matter particles to improve the small-scales of the low resolution approximate simulations, and to model the baryonic feedback on the total matter distribution. Motivated by these methods, we propose to model this displacement field by
\begin{equation}
    \label{equ:displacement1}
    \mathbf{S} = \alpha \mathbf{\nabla}
    \mathbf{\hat{O}_G}f(\delta) ,
\end{equation}
where $\alpha$ is a learnable parameter, $\delta$ is the matter overdensity as output 
by the initial layers (FastPM and LDL on dark 
matter layer), $f(\delta)$ is the source term and can be an arbitrary function of $\delta$. Here we choose it to be a power law
\begin{equation}
    \label{equ:displacement2}
    f(\delta) = (1+\delta)^{\gamma} ,
\end{equation}
with $\gamma$ a learnable parameter. $\mathbf{\hat{O}_G}$ is the Green's operator, and can be written explicitly as
\begin{equation}
    \label{equ:green}
    \mathbf{\hat{O}_G}f(\delta) = \int G(\mathbf{x-x'}) f(\delta(\mathbf{x'}))d\mathbf{x'} ,
\end{equation}
where $G(\mathbf{x-x'})$ is the Green's function and we have used $G(\mathbf{x,x'})=G(\mathbf{x-x'})$ due to translational symmetry. The convolution in above equation can be easily calculated in Fourier space as $\mathbf{\hat{O}_G}f(\delta) = G(\mathbf{k}) f(\delta)$, and we further have $G(\mathbf{k})=G(k)$ because of the rotational symmetry of the system.
Following the PGD model, we model $\mathbf{\hat{O}_G}$ in Fourier space as
\begin{equation}
    \label{equ:displacement3}
    \mathbf{\hat{O}_G} = \exp(-(k_h/k)^2)\exp(-(k/k_l)^2)\ k^n,
\end{equation}
where $k_h$, $k_l$ and $n$ are additional learnable parameters. The high pass filter $\exp(-(k_h/k)^2)$ prevents the large scale growth, since the baryonic physics that we are trying to model is an effective 
description of the 
small scale physics, while the large scales
are correctly described by the linear perturbative solution
enforced by FastPM. Together with the low pass filter $\exp(-(k/k_l)^2)$, which has the 
typical effective theory form, the operator $\mathbf{\hat{O}_G}$ is capable of learning the characteristic scale of the physics we are trying to model. Note that both the source $f(\delta)$ and the shape of $\mathbf{\hat{O}_G}$ characterizes the complex bayron and subgrid physics and cannot be derived from first principles. They can only be learned from high resolution hydrodynamical simulations.  Equation \ref{equ:displacement1} - \ref{equ:displacement3} defines the displacement field of one Lagrangian layer. We can stack multiple such layers to form a deep learning model, where each layer takes the particle output from the previous layer (which determines $\delta$ in Equation \ref{equ:displacement1}) and adds additional displacements to the particles. Such a deep model will be able to learn more complex physics. The idea is that different layers can focus on different physics components, which will differ
in terms of the scale dependence of the potential 
and its gradient, as well as in terms of the source density dependence. 

Note that if we set $\alpha=4\pi G{\bar \rho}$, $\gamma=1$, $k_h=0$, $k_l=\infty$ and $n=-2$, then Equation \ref{equ:displacement1} is exactly the gravitational force. However, unlike the true dynamics which is a second order equation in time, one for the displacement and one for the velocity/momentum, here we are trying to effectively mimic the missing baryonic physics and therefore bypass the momentum. We also allow all these parameters to vary in order to model the physics that is different from gravity. 

The final output layer is modeled as a nonlinear transformation on the particle density field:
\begin{equation}
    \label{equ:threshold}
    F(x) = \mathrm{ReLU}(b_1(1+\delta'(x))^{\mu}-b_0) ,
\end{equation}
where $F$ is the output target field, $\mathrm{ReLU}(x)$ is the 
rectified linear unit, which is zero if $x<0$
and $x$ otherwise,
$\delta'$ is the particle overdensity field after the displacement, and $b_0$, $b_1$ and $\mu$ are learnable parameters. This is motivated by the physics processes that cannot be modeled as a matter transport (i.e. displacement). In the 
example of stars, the Lagrangian displacement layers are designed to learn the effect of gas cooling and collapse. After these displacement layers, the particles are moved towards the halo center, where protogalaxies are formed and we expect star formation to happen in these dense regions. This star formation process will be modeled by Equation \ref{equ:threshold}:  the $\mathrm{ReLU}$ thresholding aims at selecting the high density regions where the star formation happens. Such thresholding is typical of a 
subgrid physics model: in the absence of this thresholding we would need to transport all of the particles 
out of the low density regions where the star 
formation does not happen, a process that 
does not have a corresponding physical model. 


\begin{table*}
\centering
\caption{The numerical parameters of LDL hybrid simulations, low resolution TNG300-3 and the target TNG300-1 hydrodynamical simulations}
\label{tab:comparison}
\begin{tabular}{P{4.0cm}|P{3.0cm}|P{3.0cm}|P{2.5cm}|P{2.5cm}}
&  FastPM + LDL & TNG300-3-Dark + LDL & TNG300-3 & TNG300-1\\
\midrule
$N_{\mathrm{particle}}$ & $625^3$ & $625^3$ & $N_{\mathrm{DM}}=625^3$ $N_{\mathrm{gas}}=625^3$ & $N_{\mathrm{DM}}=2500^3$ $N_{\mathrm{gas}}=2500^3$\\
\hline
Force / Mesh Resolution ($h^{-1} \mathrm{ckpc}$) & $164$ (FastPM)\qquad \qquad \qquad $328$ (LDL) & $4.0$ (TNG-Dark)\qquad \qquad \qquad $328$ (LDL) & $\epsilon_{\mathrm{DM, *}}=4.0$ $\epsilon_{\mathrm{gas}}=1.0$ & $\epsilon_{\mathrm{DM, *}}=1.0$ $\epsilon_{\mathrm{gas}}=0.25$ \\
\hline
Number of Time Steps / layers & $N_{\mathrm{FastPM}}=10$ $N_{\mathrm{LDL,*}}=4$ & $N_{\mathrm{TNG}}=9201$ $N_{\mathrm{LDL,*}}=3$ & $209,161$ &$6,203,062$ \\
\hline
CPU Time &$T_{\mathrm{IC}}=2.3\ \mathrm{h}$  $T_{\mathrm{FastPM}}=5.1\ \mathrm{h}$ $T_{\mathrm{LDL,*}}=0.4\ \mathrm{h}$  & $T_{\mathrm{TNG}}=5.9\ \mathrm{kh}$ $T_{\mathrm{LDL,*}}=0.3\ \mathrm{h}$ & $0.05\ \mathrm{Mh}$ & $34.9\ \mathrm{Mh}$\\
\bottomrule
\end{tabular}
\begin{tablenotes}
\small
\item The LDL parameters for generating stellar mass. The architecture for other observables can be found in Table \ref{tab:architecture}. The total CPU time for LDL is 7.8 hours, compared to $5 \times 10^4$ for the full hydro TNG300-3. Despite this the LDL outperforms the full hydro at the same resolution in all of the outputs. In this paper we are primarily concerned with a proof of principle and both FastPM and LDL are run with Python. We expect the CPU time to be further reduced if running them with C. 
\end{tablenotes}
\end{table*}

In this work we use both FastPM and N-body simulations, combining them with LDL to predict the baryon observables from the linear density map. 
We consider modeling the stellar mass, kSZ signal, tSZ signal and X-ray at redshift $z=1$, $z=0.5$ and $z=0$. The dark matter particles are firstly evolved to these redshifts with FastPM, and then passed to the LDL networks for modeling the baryons. The parameters in LDL are optimized by matching the output with the target fields from TNG300-1 hydrodynamical simulation \cite{pillepich2018first, naiman2018first, marinacci2018first, springel2018first, nelson2018first}. Since the kSZ signal is proportional to the electron momentum $n_ev_z$, the tSZ signal is proportional to the electron pressure $n_eT$, and the X-ray emissivity is approximately proportional to $n_e^2T^{0.5}$ (we only consider the bremsstrahlung effect and ignore the Gaunt factor), we will model these fields in the rest of this paper.

Apart from FastPM, we also consider combining LDL models with full N-body simulations. We take the particle data at redshift $z=1$, $z=0.5$ and $z=0$ from TNG300-3-Dark, a low resolution dark-matter-only run of the TNG300 series, and feed the particles to LDL models. In the next section we will compare the performance of these two hybrid simulations against the target high resolution hydrodynamical simulation.

We summarize the numerical parameters of these simulations in Table \ref{tab:comparison}. We also list TNG300-3, the low resolution hydrodynamic run of TNG300. TNG300-3 has the resolution of our hybrid simulations, and is a natural reference to compare the performance of our models with. Note that the mass resolution, force / mesh resolution and time resolution of our hybrid simulations are significantly lower than the target simulation, and the N-body simulation and deep learning networks are also much cheaper to run compared to simulating hydrodynamics. As a result, the FastPM-based and N-body-based hybrid simulations are 7 and 4 orders of magnitudes cheaper than the target simulation, respectively. When comparing to TNG300-3, our hybrid simulations are still 4 and 1 orders of magnitudes cheaper, respectively, and 
we show that by being trained on the high resolution TNG300-1 our 
simulations are superior to TNG300-3, and comparable to TNG300-1. 

\section*{Results}

\begin{figure}
\centering
\includegraphics[height=0.878\textheight]{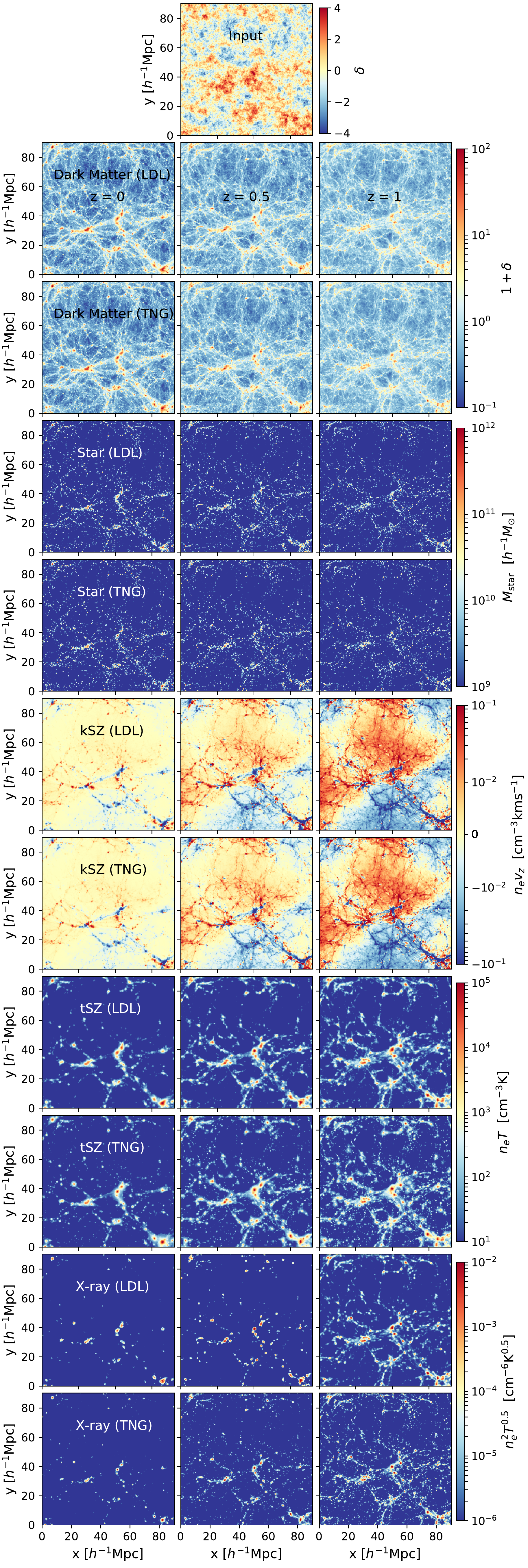}
\caption{Visualization of slices of the simulations: the first row is the input linear density field. The 2nd, 4th, 6th, 8th and 10th rows are predictions of dark matter overdensity, stellar mass, electron momentum density $n_ev_z$ (kSZ signal), electron pressure $n_eT$ (tSZ signal) and $n_e^2T^{0.5}$ (X-ray signal) from our FastPM+LDL hybrid simulation, respectively. The 3rd, 5th, 7th, 9th and 11th rows are the corresponding target fields from TNG300-1 hydrodynamical simulation. The left panel, middel panel and right panel are from redshift $z=0$, $z=0.5$ and $z=1$, respectively. The slices are from a $90.2\times 90.2\times 32.8 h^{-1}\mathrm{Mpc}$ sub-box of the test set.}
\label{fig:visual}
\end{figure}

We show in Figure \ref{fig:visual} the visualization of slices of the input linear density field and the output dark matter of our FastPM-based hybrid simulation, as well as the target fields in hydrodynamical simulation. Visual 
agreement between the two is very good. The 
results are shown for the dark matter density, stellar mass density, electron momentum density $n_ev_z$, where $n_e$ is electron density and 
$v_z$ radial velocity, 
electron 
pressure $n_eT$, where $T$ is the gas temperature, and X-ray emission proportional 
to $n_e^2T^{0.5}$. 

\subsection*{Power Spectrum}

\begin{figure}
\centering
\includegraphics[width=\linewidth]{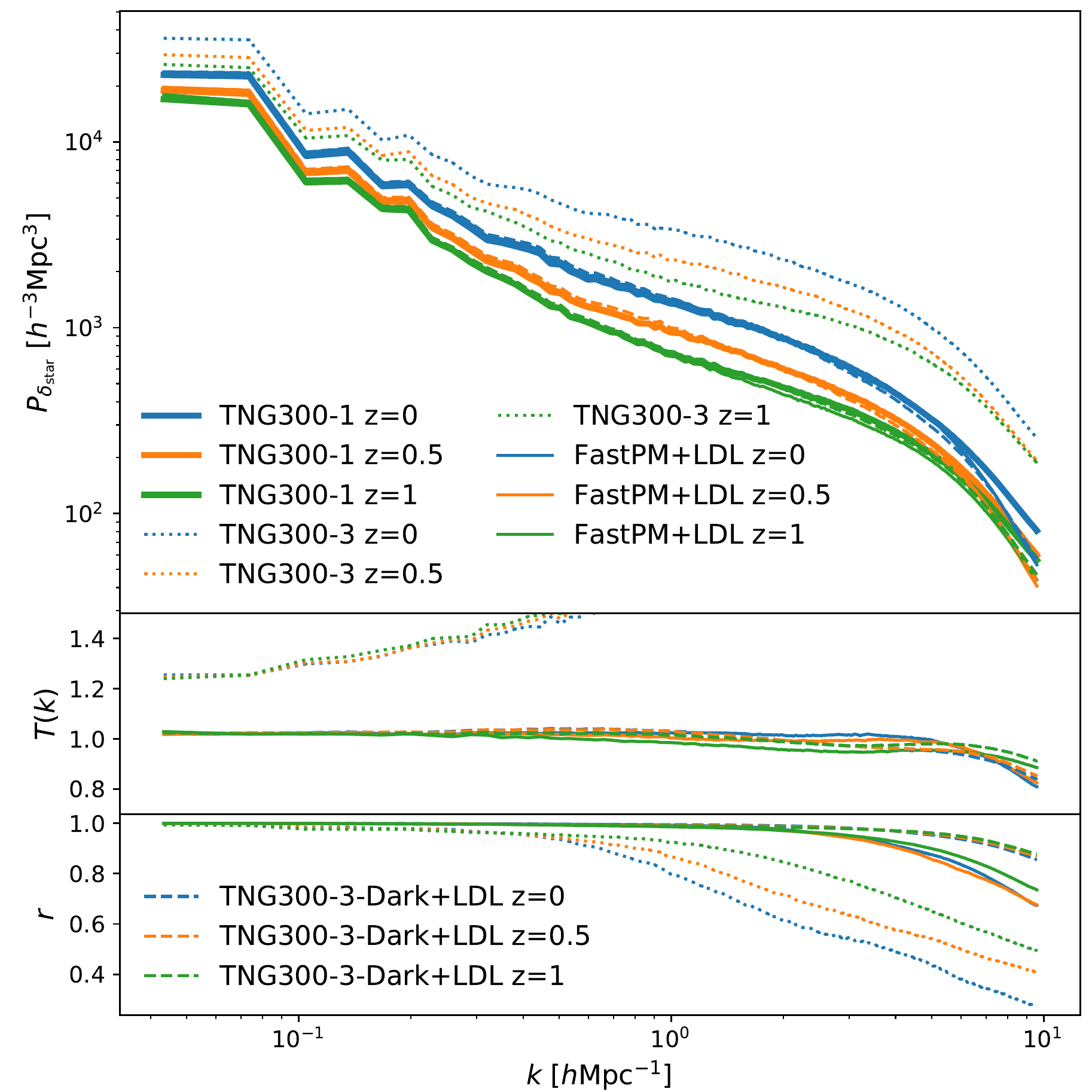}
\caption{Comparison of the 3D power spectrum (top panel), transfer function (middle panel) and cross correlation coefficient (bottom panel) of the stellar mass overdensity. We compare the LDL hybrid simulations, TNG300-3 and the target TNG300-1 hydrodynamical simulation, at redshifts 0, 0.05 and 1.}
\label{fig:PS_star}
\end{figure}

\begin{figure}
\centering
\includegraphics[width=\linewidth]{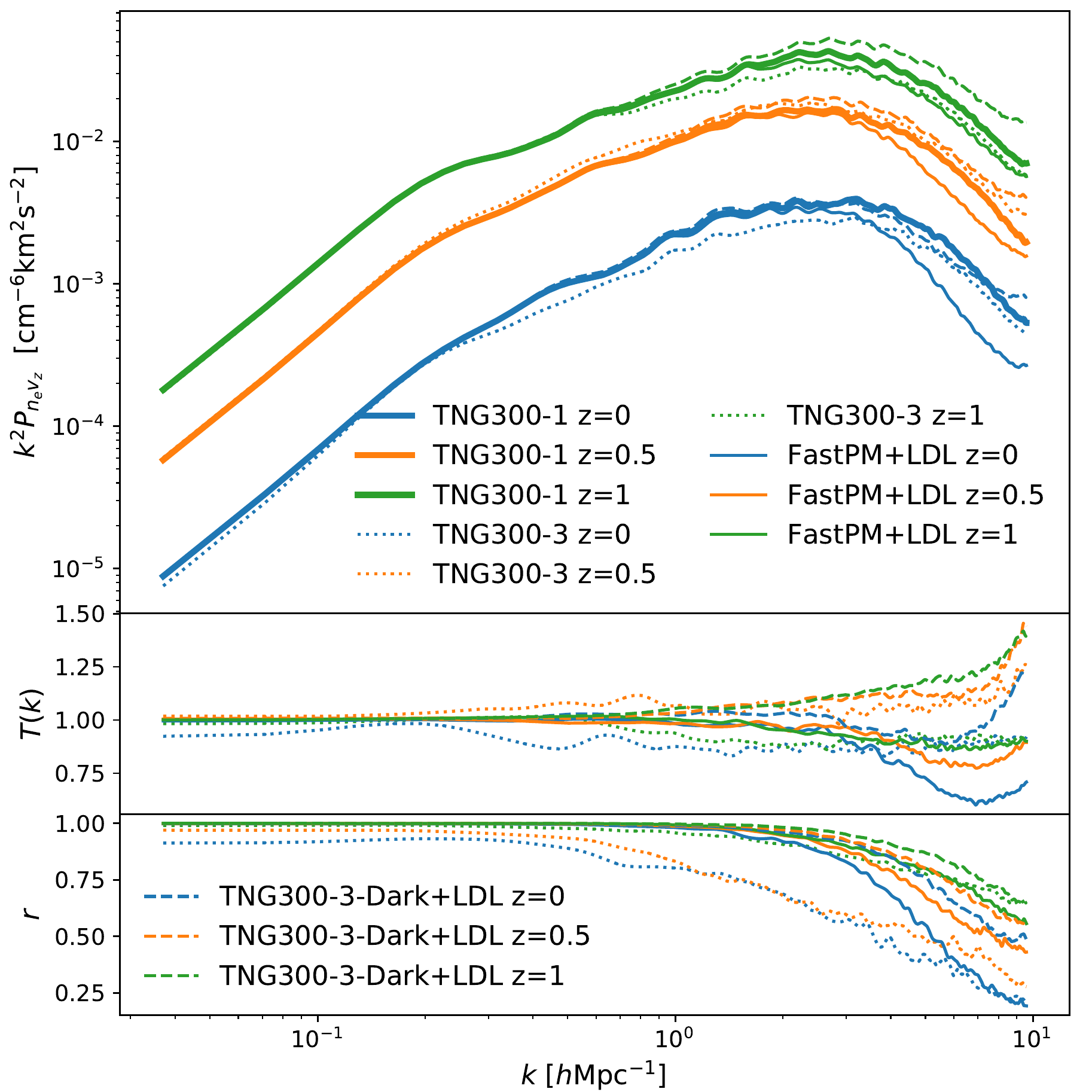}
\caption{Comparison of the 2D power spectrum (top panel), transfer function (middle panel) and cross correlation coefficient (bottom panel) of the electron momentum density $n_ev_z$ (proportional to kSZ signal) between the LDL hybrid simulations, TNG300-3 and the target TNG300-1 hydrodynamical simulation.}
\label{fig:PS_kSZ}
\end{figure}

\begin{figure}
\centering
\includegraphics[width=\linewidth]{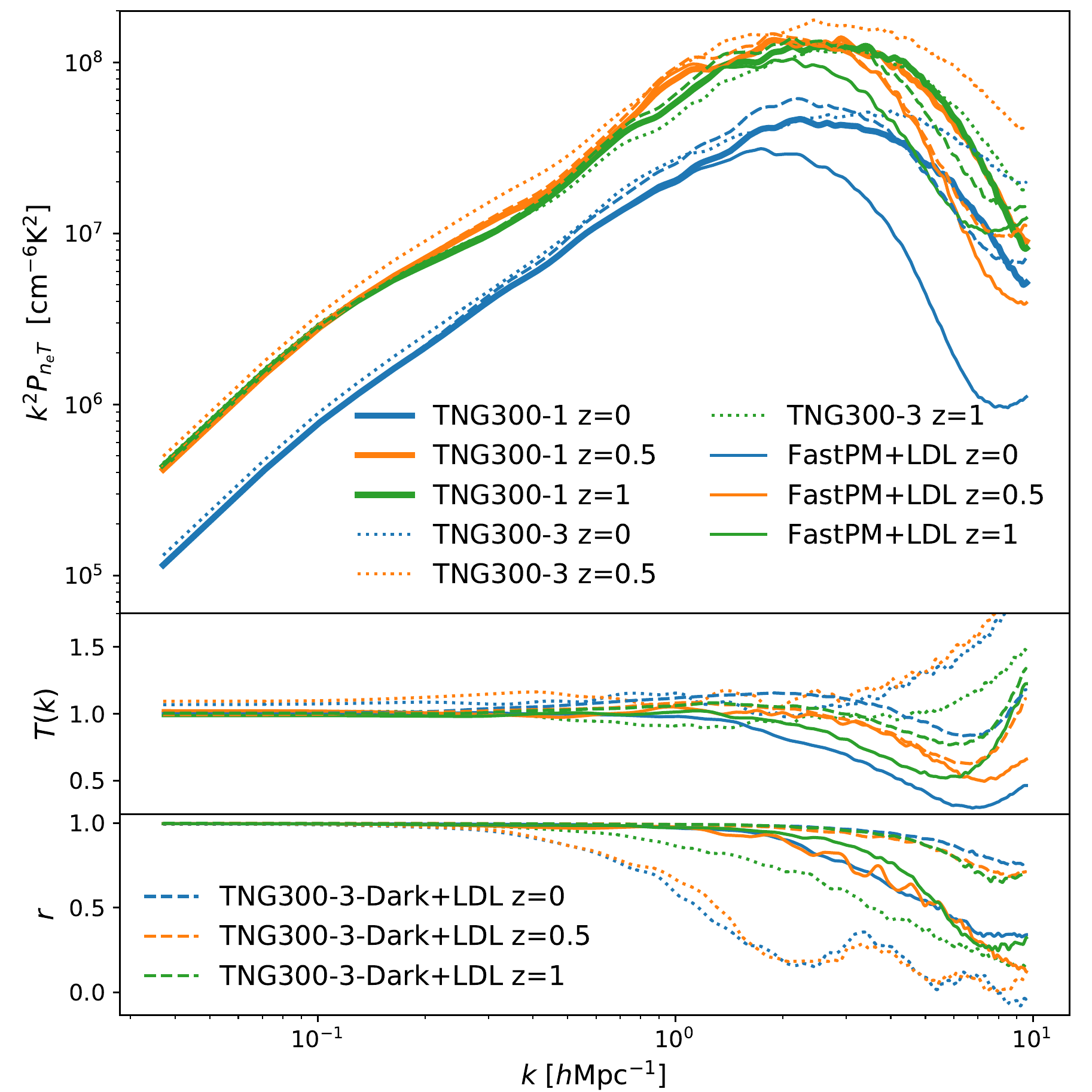}
\caption{Comparison of the 2D power spectrum (top panel), transfer function (middle panel) and cross correlation coefficient (bottom panel) of the electron pressure $n_eT$ (proportional to tSZ signal) between the LDL hybrid simulations, TNG300-3 and the target TNG300-1 hydrodynamical simulation.}
\label{fig:PS_tSZ}
\end{figure}

\begin{figure}
\centering
\includegraphics[width=\linewidth]{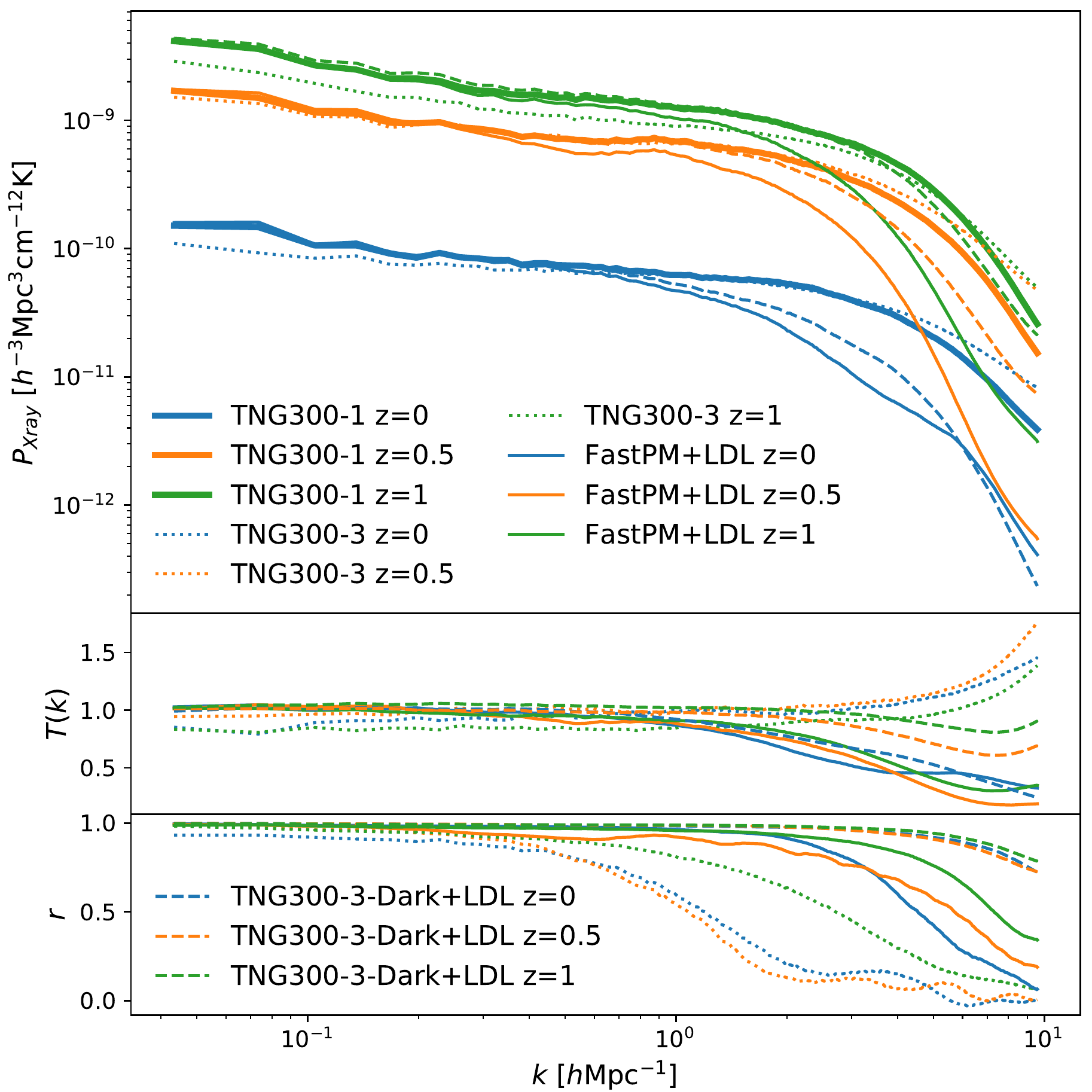}
\caption{Comparison of the 3D power spectrum (top panel), transfer function (middle panel) and cross correlation coefficient (bottom panel) of the gas property $n_e^2T^{0.5}$ (proportional to X-ray emissivity) between the LDL hybrid simulations, TNG300-3 and the target TNG300-1 hydrodynamical simulation.}
\label{fig:PS_Xray}
\end{figure}

We measure the summary statistics of these fields and compare them quantitatively.
We firstly compare the power spectrum, the most widely used summary statistics in cosmology. We define the transfer function as:
\begin{equation}
    \label{equ:T}
    T(k) = \sqrt{\frac{P_{\mathrm{predict}}(k)}{P_{\mathrm{target}}(k)}},
\end{equation}
and the cross correlation coefficient as:
\begin{equation}
    \label{equ:r}
    r(k) =  \frac{P_{\mathrm{predict,target}}(k)}{\sqrt{P_{\mathrm{predict}}(k)P_{\mathrm{target}}(k)}},
\end{equation}
where $P_{\mathrm{predict,target}}(k)$ is the cross power spectrum between the predicted field and the target field. In this paper we use the whole $205h^{-1}\mathrm{Mpc}$ box for the measurement of power spectrum and cross correlations in order to compare the large scale modes. We show the 3D or 2D  power spectrum, transfer function and cross correlation coefficient of the stellar mass overdensity $\delta_*$, electron momentum $n_ev_z$, electron pressure $n_eT$ and X-ray intensity $n_e^2T^{0.5}$ in Figures \ref{fig:PS_star}, \ref{fig:PS_kSZ}, \ref{fig:PS_tSZ} and \ref{fig:PS_Xray}, respectively. On large scale and intermediate scale our hybrid simulations match well with the target fields, while TNG300-3 agreement is worse, especially for the stellar mass. The large bias of TNG300-3 stellar mass might be partially due to the fact that the low resolution TNG300-3 cannot resolve the stars in small halos. 
In contrast, by training on high resolution hydro simulations TNG300-1, our low resolution hybrid simulations are able to model those small galaxies better than the full hydro simulation at the 
same resolution. 

On the small scales all of the predicted fields show some deviations from the targets. We discuss possible reasons for these in the next Section. We also see that the full-N-body-based hybrid simulation normally predicts larger small scale power than the FastPM-based simulation. This is likely due to the fact that the 10-layer FastPM cannot fully model the small halos and halo internal structures, and its simulated dark matter distribution is less clustered on small scale compared to full N-body simulations, making its predicted baryon fields less clustered. Overall, the predicted power spectrum from the N-body-based hybrid simulation is better, although it can predict too much small scale power (e.g. the kSZ signal at redshift 1). 

The cross correlation coefficients are also shown in these 
Figures. We observe that the hybrid simulations are significantly better than those of TNG300-3, with the N-body-based hybrid simulation a bit higher than the FastPM-based simulation. Note that in principle the cross correlation coefficient, which quantifies the agreement of phases of Fourier modes, is a more important statistics than the transfer function, because the transfer function can always be corrected to unity by multiplying the predicted fields with the reciprocal of the transfer function. This again suggests that the baryon maps of our models are closer to the ground truth than full hydrodynamical simulations at the same resolution.

\subsection*{Cross Correlations between different tracers}

\begin{figure*}
\centering
\includegraphics[width=\linewidth]{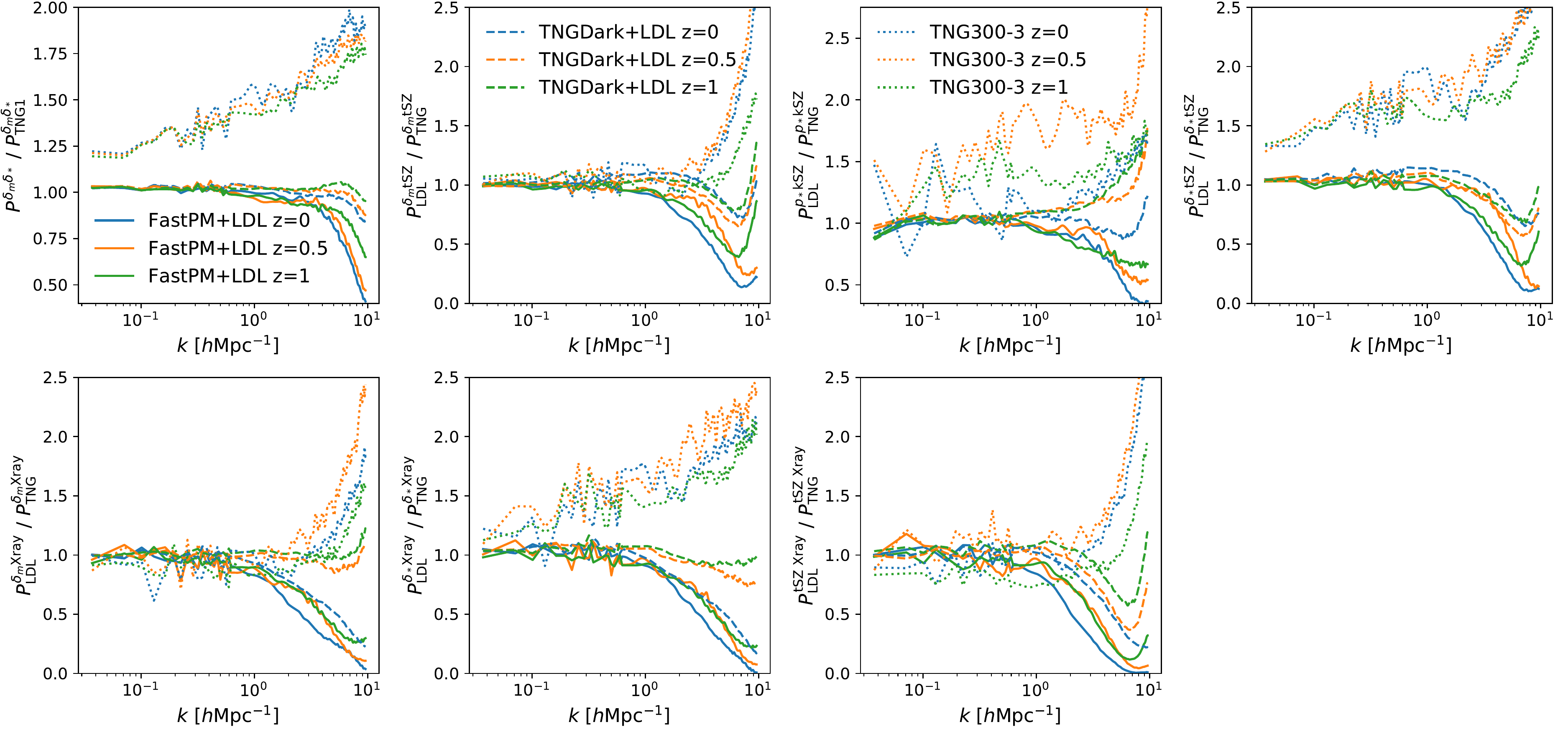}
\caption{The ratio of the 2D cross power spectrum of different observables between the LDL hybrid simulations and the target TNG300-1 hydrodynamical simulation. The first row show the cross power spectrum of matter and stellar mass (1st panel), matter and tSZ signal (2nd panel), stellar momentum density and kSZ signal (3rd panel), and stellar mass and tSZ signal (4th panel). The second row shows the cross power spectrum of matter and X-ray (1st panel), stellar mass and X-ray (2nd panel), and tSZ and X-ray (3rd panel).}
\label{fig:CrossPS}
\end{figure*}

Probes of the large-scale structure, such as weak lensing, galaxy survey and clusters, are strongly correlated because they are all determined by the same underlying matter distribution. There is additional information in the cross correlations between these probes which cannot be obtained by analyzing each observable independently. The cross correlation also has the advantage that the noise
does not add to it. 
Our hybrid simulation is able to generate various observables simultaneously with a low computational cost, so it is potentially promising for cross correlation analysis. 
Here we investigate the predicted cross correlations between weak lensing convergence, mass weighted galaxies, tSZ and X-ray, as well as the cross correlation between the galaxy momentum and kSZ signal. We show the ratio of the predicted cross power spectrum and the TNG300-1 in Figure \ref{fig:CrossPS}. Similar to the auto power spectrum analysis, our predicted cross power spectrum is consistent with the target simulation on large scale, while TNG300-3 does not agree that well. On small scales FastPM-based hybrid simulation tends to underestimates the power, while TNG300-3 tends to overestimate the power. One can compare the second panel of Figure \ref{fig:CrossPS} (cross power spectrum between the matter and tSZ) with Figure 2 of \cite{troster2019painting}, where GAN and VAE 
is used to predict the gas pressure from N-body simulations. We observe that the deviation of full-N-body-based hybrid simulation is comparable to the deviations of GAN and VAE. We note that for the standard 
deep learning architectures employed by 
GAN or VAE the number of parameters being fit 
is very large, in contrast to our approach. 


\section*{Discussion}

We propose a novel Lagrangian Deep Learning (LDL) model for learning the effective 
physical laws from the outputs of either 
simulations or real data. Specifically, in this paper we focus on learning the physics that controls baryon 
hydrodynamics in the cosmological simulations. We build hybrid simulations by combining N-body / quasi N-body gravity solver with LDL models. We show that both the FastPM-based and N-body-based hybrid simulations are able to generate maps of stellar mass, kSZ, tSZ and X-ray of various redshifts from the linear density field, and their computational costs are 7 and 4 orders of magnitudes lower than the target high resolution hydrodynamical simulation. We perform the auto power spectrum analysis and the cross correlation analysis among these fields, and we show that they outperform the hydrodynamical simulation at the same resolution. 

The LDL model is motivated by the desire to provide an effective description of the underlying physics. Such a description 
must obey all the symmetries of the problem, and  rotation and translation invariance are the two key 
symmetries, but other symmetries of the 
problem such as mass conservation 
may also appear. In this paper we argue that implementing these 
symmetries creates a generative model that 
is learning an effective description of the physical laws as opposed to learning the 
data distribution. This is because the 
symmetries are the only constraints on 
the generative model that 
must be implemented explicitly, everything 
else can be learned from the data. 
Here we propose that the learning of the 
generative model 
can be implemented by composing layers of
displacements acting on the effective 
particles describing the physical properties
of a system such as a fluid, 
moving the particles following the Lagrangian approach. The displacement of the particles 
can be understood as a result of the 
underlying physical processes, with particle transport a consequence of processes such as gas cooling and heating, feedback, turbulence etc. The output layer is a nonlinear transformation with thresholding on the particle density field, which models physics processes such as star formation.

Translational and rotational symmetry of the system put strong constraints on the model and therefore the Green's operator can be written as a function in Fourier Space that only depends on the amplitude of $k$.
This allows us to use very few parameters to model the complex processes and produce maps of observables.
Thus even
though we want to describe systems of extremely 
high dimensionality ($10^8$ or more), 
the underlying effective 
physics description requires a handful of 
parameters only. 

The small number of free parameters also make the model stable and easy to train. An important advantage is that 
we can use the small number of parameters as 
effective physics description of a
complicated microphysics model, similar to the 
free parameters that arise from 
renormalization in the effective 
field theory descriptions of microphysics. This 
suggests that our LDL approach can replace 
other effective descriptions used to model 
the process of star formation. 
In cosmology such simplified models are often based on 
first identifying the dark matter halos in a 
dark matter simulation only, 
followed by some effective description of 
how to populate these halos with stars. 
Compared to such semi-analytical approaches which often rely on non-differentiable models, 
our approach is explicitly differentiable, 
such that we can use backpropagation to 
derive a gradient of the final observables 
with respect to the initial density field. 
This can be easily embedded into the forward modeling framework to reconstruct the 
initial conditions from the observations \cite{seljak2017towards}. 

Our current implementation outperforms the full 
hydro simulation at the same resolution, but 
does not match perfectly the higher 
resolution hydro simulation. LDL
deviates from the 
full simulation results mostly on small scales. This is 
expected, since the factor of 64 lower
mass resolution means there is some
information in the full simulation that 
cannot be recovered. 
Specifically, 
we use a low resolution mesh for calculating the displacements in the LDL model (cell size $0.328h^{-1}\mathrm{Mpc}$, see Table \ref{tab:comparison}). The low resolution mesh limits the ability of LDL to model the small scale baryon distribution.
Moreover, to ensure the correct large scale distribution, we apply a smoothing operator (Equation \ref{equ:smth}) to the fields before calculating the loss function, which downweights the small scale contribution to the loss function. 

LDL trains on hydrodynamic 
simulations and is not meant to 
replace but to complement them: for example, 
it can interpolate a coarse grid and scale them to 
larger volumes and higher resolutions. 
In contrast, LDL has the 
potential to eliminate the need for the 
semi-analytic methods, which are the 
current standard paradigm in the large 
scale structure. These methods run N-body 
simulations first and then 
populate their dark matter halos 
using a semi-analytic 
prescription for the observable. 
LDL can not only achieve results that are on 
par with the full hydro at the same 
resolution, which is superior to these 
semi-analytic approaches, it also achieves 
this with of order 10 time steps, in
addition to up to 6 LDL layers, 
in contrast to $10^3$ in an N-body simulations.
We expect this will lead both to development of 
realistic simulations that cover the full 
volume of the cosmological LSS surveys, and 
to analysis of these LSS surveys with LDL
effective parameters as the nuisance 
parameters describing the astrophysics of the
galaxy formation. 


\subsection*{Dataset}
IllustrisTNG is a suite of cosmological magneto-hydrodynamical simulations of galaxy formation and evolution \cite{pillepich2018first, naiman2018first, marinacci2018first, springel2018first, nelson2018first}. It consists of three runs of different volumes and resolutions: TNG50, TNG100 and TNG300 with sidelengths of $35h^{-1}\mathrm{Mpc}\approx50\mathrm{Mpc}$, $75h^{-1}\mathrm{Mpc}\approx100\mathrm{Mpc}$ and $205h^{-1}\mathrm{Mpc}\approx300\mathrm{Mpc}$, respectively. IllustrisTNG follows the evolution of the dark matter, gas, stars and supermassive black holes, with a full set of physical models including star formation and evolution, supernova feedback with galactic wind, primordial and metal-line gas cooling, chemical enrichment, black hole formation, growth and multi-mode feedback. The IllustrisTNG series evolves over a redshift range $z=127$ to the present $z=0$ in a $\mathrm{\Lambda CDM}$ cosmology, with parameters $\Omega_m=0.3089$, $\Omega_b=0.0486$, $\Omega_{\Lambda}=0.6911$, $H_0=67.74\mathrm{km\ s^{-1}\ Mpc^{-1}}$, $\sigma_8=0.8159$ and $n_s=0.9667$. 

In this paper we train our models against TNG300-1, the highest resolution of the TNG300 run. TNG300-1 evolves $2500^3$ dark matter particles and an initial number of $2500^3$ gas cells, with a comoving force resolution $\epsilon_{\mathrm{DM,stars}}=1.0h^{-1}\mathrm{kpc}$, $\epsilon_{\mathrm{gas,min}}=0.25h^{-1}\mathrm{kpc}$ and $\epsilon_{\mathrm{BH,max}}=5.84h^{-1}\mathrm{kpc}$. The dark matter mass resolution is $4.0\times 10^7 h^{-1}M_{\odot}$, and the target baryon mass resolution is $7.6\times 10^6  h^{-1}M_{\odot}$ (see Table \ref{tab:comparison}). 

We also compare the model performance with TNG300-3, the hydro run with the same resolution as our hybrid simulations. The mass resolution and force resolution of TNG300-3 are 64 and 4 times lower than TNG300-1, 
respectively.

\subsection*{Details of the Hybrid Simulation}

The 10-step FastPM is run in a $205 h^{-1}\mathrm{Mpc}$ periodic box, but with only $N=625^3$ particles and force resolution $B=2$. We generate the initial condition at redshift $z=9$ using second order Lagrangian perturbation theory (2LPT), with the same random seed and linear power spectrum as Illustris-TNG. The linear density map is generated with a $N=1250^3$ mesh to improve the accuracy on small scale \cite{dai2020high}. The box is then evolved to redshift $0$ with 10 time steps that are linearly separated in scale factor $a$. Three snapshots are produced at redshift $z=0$, $0.5$ and $1$, which are passed to LDL for generating maps of baryonic observables at these redshifts. Note that our mass, force and time resolutions are 64, 164 and 620,000 times lower than the target simulation TNG300-1, respectively.

Instead of running 10-step FastPM, we also tried using the particle data from the full N-body simulation TNG300-3-Dark. TNG300-3-Dark is the dark-matter-only run of the low resolution TNG300-3. It includes $N=625^3$ dark matter particles (same as our FastPM setup), but the force and time resolution is significantly higher. A detailed comparison between FastPM, TNG300-3-Dark and TNG300-1 can be found in Table \ref{tab:comparison}.

\begin{table}
\centering
\caption{The LDL architecture for predicting different baryon observables}
\label{tab:architecture}
\begin{tabular}{P{3.05cm}|P{0.8cm}|P{0.2cm}|P{0.2cm}|P{0.2cm}|P{0.2cm}|P{0.2cm}|P{0.2cm}}
& stellar mass & \multicolumn{2}{c|}{kSZ} & \multicolumn{2}{c|}{tSZ}  & \multicolumn{2}{c}{X-ray} \\
& & $n_e$ & $v_z$ & $n_e$ & $T$ & $n_e$ & $T$\\
\midrule
Displacement Layer  (Eq. \ref{equ:displacement1}) & 2 & 1 & 0 & 1 & 2 & 2 & 2\\
\hline
Output Layer (Eq. \ref{equ:threshold}) & 1 & 1 & 0 & 1 & 1 & 1 & 1\\
\hline
Total number of layers & 3 (4) & \multicolumn{2}{c|}{2 (3)} & \multicolumn{2}{c|}{5 (6)} & \multicolumn{2}{c}{6 (7)} \\
\hline
Total number of free parameters & 13 (18) & \multicolumn{2}{c|}{10 (13)} & \multicolumn{2}{c|}{21 (26)} & \multicolumn{2}{c}{26 (31)} \\
\bottomrule
\end{tabular}
\begin{tablenotes}
\small
\item For FastPM-based hybrid simulation, we add one more displacement layer to improve the small scale dark matter distribution. The corresponding $N_{\mathrm{layer}}$ and $N_{\mathrm{parameter}}$ are shown in parentheses.
\end{tablenotes}
\end{table}

The details of the LDL model are described in the 
main text. 
We use a $N=625^3$ mesh for calculating the displacement and generating the hydro maps. The architecture of the model is shown in Table \ref{tab:architecture}. Specifically, for FastPM input, we firstly add a Lagrangian displacement layer and the output is matched to the density field of the full N-body simulation TNG300-3-Dark. This layer is intended to improve the small scale structure of FastPM and is shared by all hydro outputs (we do not add this layer for TNG300-3-Dark input). Then for different observables, we train different displacement layers and output layer: 1. For stellar mass, we add two displacement layer to mimic gas cooling and collapse, and one output layer to model star formation. 2. For kSZ signal, we use one displacement layer and one output layer to model the electron number density field. We assume that the velocities of gas trace dark matter, so the velocity field can be directly estimated from the dark matter particles: $v(x) = \frac{p(x)}{\rho(x)}$, where $p(x)$ is the momentum density field and $\rho(x)$ is the matter density field. The kSZ map is obtained by multiplying the electron density field and the velocity field. 3. For tSZ signal map, we generate the electron number density field with one displacement layer and one output layer, and generate the gas temperature map with two displacement layers and one output layer. Then the two fields are multiplied to produce the tSZ signal. 4. The modeling of X-ray is similar to tSZ, except that now we use two displacement layer to model the electron density.

\subsection*{Model Training and Loss Function}
As described above, the output of the LDL model is a $N=625^3$ mesh. We retain $77.7\%$ of the pixels for training, $13.8\%$ for validation and $8.5\%$ for test. Similar to \cite{zhang2019from}, we split between training, validation, and test set following a ``global'' cut. The test set forms a sub-box of $90.2h^{-1}\mathrm{Mpc}$ per side, and the validation set is a $90.2 \times 114.8 \times 114.8\  h^{-1}\mathrm{Mpc}$ sub-box. The rest of the $205 h^{-1}\mathrm{Mpc}$ box is used for training.

For stellar mass and the electron number density field in kSZ map, we define the loss function as:
\begin{equation}
    \label{equ:loss1}
    \mathcal{L} = \sum_{i=1}^{N}\|\mathbf{\hat{O}_s}F_{\mathrm{LDL}}(x_i)-\mathbf{\hat{O}_s}F_{\mathrm{TNG}}(x_i)\| ,
\end{equation}
where $\|$ is $L_2$ norm, $i$ labels the mesh cell, $F_{\mathrm{LDL}}(x)$ is the generated map from LDL, $F_{\mathrm{TNG}}(x)$ is the true hydro map from IllustrisTNG, and $\mathbf{\hat{O}_s}$ is a smoothing operator defined in Fourier space:
\begin{equation}
    \label{equ:smth}
    \mathbf{\hat{O}_s} = 1 + (\frac{k}{1h\mathrm{Mpc^{-1}}})^{-n} .
\end{equation}
Here $n$ is a hyperparameter that determines the relative weight between the large scale modes and the small scale modes. Without the $\mathbf{\hat{O}_s}$ operator, the model focuses on the small scale distribution and results in a biased large scale power due to the small number of large scale modes
relative to small scale modes. We apply $\mathbf{\hat{O}_s}$ operator to put more weight on the large scale distribution. For most of the baryon maps we set $n=1$, except for the electron number density $n_e$ of the X-ray map we set it to be $n=0.85$.

For the tSZ map, we use a different loss function to improve the performance. We firstly train the electron density map with the following loss function:
\begin{equation}
    \label{equ:losstSZ1}
    \mathcal{L}_{n_e}^{\mathrm{tSZ}} = \sum_{i=1}^{N} \| \mathbf{\hat{O}_s} [n_{e_{\mathrm{LDL}}}(x_i)T_{\mathrm{TNG}}(x_i)] - \mathbf{\hat{O}_s}[n_{e_{\mathrm{TNG}}}(x_i)T_{\mathrm{TNG}}(x_i)] \| ,
\end{equation}
where $n_{e_{\mathrm{LDL}}}(x)$ is the learned electron number density map, $n_{e_{\mathrm{TNG}}}(x)$ is the true electron number density map, and $T_{\mathrm{TNG}}$ is the true temperature map. This 
means we multiply the electron number density field with the temperature field before calculating the loss function. This procedure puts more weight on the large clusters and improves the quality of the generated tSZ maps. Note that this electron density field is different from the electron density field for predicting the kSZ signal. Similarly, after we obtain the learned electron number density field $n_{e_{\mathrm{LDL}}}(x)$, we train the temperature map with the following the loss function:
\begin{equation}
    \label{equ:losstSZ2}
    \mathcal{L}_{T}^{\mathrm{tSZ}} = \sum_{i=1}^{N} \| \mathbf{\hat{O}_s} [n_{e_{\mathrm{LDL}}}(x_i)T_{\mathrm{LDL}}(x_i)] - \mathbf{\hat{O}_s}[n_{e_{\mathrm{TNG}}}(x_i)T_{\mathrm{TNG}}(x_i)] \| .
\end{equation}
Here $n_{e_{\mathrm{LDL}}}(x)$ is the electron density field we just learned and is fixed, and $T_{\mathrm{LDL}}(x)$ is the target temperature field that we are trying to optimize.

For the X-ray map, similar to the tSZ signal, we train the electron density and gas temperature maps successively with the following loss functions:

\begin{equation}
    \label{equ:lossXray1}
    \mathcal{L}_{n_e}^{\mathrm{X}} = \sum_{i=1}^{N} \| \mathbf{\hat{O}_s} [n^2_{e_{\mathrm{LDL}}}(x_i)T^{0.5}_{\mathrm{TNG}}(x_i)] - \mathbf{\hat{O}_s}[n^2_{e_{\mathrm{TNG}}}(x_i)T^{0.5}_{\mathrm{TNG}}(x_i)] \| ,
\end{equation}
\begin{equation}
    \label{equ:lossXray2}
    \mathcal{L}_{T}^{\mathrm{X}} = \sum_{i=1}^{N} \| \mathbf{\hat{O}_s} [n^2_{e_{\mathrm{LDL}}}(x_i)T^{0.5}_{\mathrm{LDL}}(x_i)] - \mathbf{\hat{O}_s}[n^2_{e_{\mathrm{TNG}}}(x_i)T^{0.5}_{\mathrm{TNG}}(x_i)] \| .
\end{equation}

Again, the electron number density field and gas temperature field for X-ray are different from the fields used for generating kSZ and tSZ.

Because the number of free parameters is relatively small, in this work we use the L-BFGS-B algorithm \cite{byrd1995limited} for optimizing the model parameters.

\section*{Acknowledgements}
We thank Dylan Nelson and the IllustrisTNG team for kindly providing the linear power spectrum, random seed and the numerical parameters of the IllustrisTNG simulations. We thank Yu Feng for helpful discussions. The majority of the computation were performed on NERSC computing facilities Cori, billed under the cosmosim and m3058 repository.  National Energy Research Scientific Computing Center (NERSC) is a U.S. Department of Energy Office of Science User Facility operated under Contract No. DE-AC02-05CH11231. The power spectrum analysis in this work is performed using the open-source toolkit nbodykit \cite{hand2018nbodykit}.

\normalsize
\bibliography{references}


\end{document}